\documentclass[preprint,showpacs,11pt]{revtex4}

\usepackage{graphicx}
\usepackage{dcolumn}
\usepackage{bm}

\begin{document}


\title{Semiclassical Methods for Hawking Radiation from a Vaidya Black Hole\footnote{Some findings in this work had already been worked out in \cite{26}.\\ Accepted for publication in \textit{Int. J. Mod. Phys. \textbf{A}}}}

\author{Haryanto M. Siahaan}%
 \email{anto_102@students.itb.ac.id}
\affiliation{
\small{Theoretical Physics Laboratory, THEPI Division\\
Faculty of Mathematics and Natural Sciences\\ Institut Teknologi Bandung\\
Jalan Ganesha 10, Bandung 40132, Indonesia\\
and\\
St. Aloysius School\\ Jalan Batununggal Indah II/30\\ Bandung 40266, Indonesia}}

\author{Triyanta}
 \email{triyanta@fi.itb.ac.id}
\affiliation{
\small{Theoretical Physics Laboratory, THEPI Division\\
and\\
INDONESIA Center for Theoretical and Mathematical Physics (ICTMP)\\
Faculty of Mathematics and Natural Sciences\\ Institut Teknologi Bandung\\
Jalan Ganesha 10, Bandung 40132, Indonesia}}


\begin{abstract}
We derive the general form of Hawking temperature for Vaidya black hole in the tunneling pictures. This kind of black hole is regarded as the description of a more realistic one since it's time dependent decreasing mass due to the evaporation process. Clearly, the temperature would be time dependent as our findings. We use the semiclassical methods, namely radial null geodesic and complex paths methods. Both methods are found to give the same results. Then, we discuss the possible form of corresponding entropy.
\end{abstract}

\maketitle

\section{Introduction}
In 1974, Hawking startled the physics community by proving that black hole evaporates particles \cite{1,2}. It contradicts with classical general relativistic definition of a black hole, an object that nothing can escape from it \cite{3}. Hawking's derivation was a quantum field theoretically. With Hartle \cite{22}, he also derived the black hole temperature by using Feynman path integral which can be categorized as semiclassical method, but it still has such mathematical complexity. Semiclassical methods for black hole radiation \cite{4,7} were developed in the past decade and they attract many attentions, e.g. Refs. \onlinecite{15} and \onlinecite{16}.\newline
There are two ways to perform semi-classical analysis for a black hole radiation. The first is by the use of radial null geodesic method developed by Parikh and Wilczek \cite{4}. In this method, one has to get the expression dr/dt from the radial null geodesic condition, $
ds^2  = d\Omega  = 0$, for a metric that has the form $ds^2  =  - a\left( r \right)dt^2  + b\left( r \right)dr^2  + r^2 d\Omega ^2 
$. Then, the obtained expression is used to calculate the imaginary part of the action for the process of s-wave emission and relates it to the Boltzman factor for emission to get Hawking temperature.\newline
The second one is called complex paths method was developed by Padmanabhan et.al. \cite{6,7}. In the method, the scalar wave function is determined by the ansatz $\phi \left( {r,t} \right) = \exp \left[ {{{ - iS\left( {r,t} \right)} \mathord{\left/
 {\vphantom {{ - iS\left( {r,t} \right)} \hbar }} \right.
 \kern-\nulldelimiterspace} \hbar }} \right]
$ where $S\left( {r,t} \right)
$ is the action for a single scalar particle. Inserting this ansatz into the Klein-Gordon equation in a gravitational background, one yields an equation for the action $S\left( {r,t} \right)
$ which can be solved by the Hamilton-Jacobi method. After obtaining the action, one can get the probability for outgoing and ingoing particles, $P\left[ {{\rm{out}}} \right] = \left| {\phi _{{\rm{out}}} } \right|^2 
$ and $P\left[ {{\rm{in}}} \right] = \left| {\phi _{{\rm{in}}} } \right|^2 
$, respectively. The Hawking temperature can be obtained by using the 'principle of detailed balance' \cite{6,7}, $P\left[ {{\rm{out}}} \right] = \exp \left[ { - \beta E} \right]P\left[ {{\rm{in}}} \right] = \exp \left[ { - \beta E} \right]
$, since all particles must be absorbed by the black hole.\newline
In this paper, we consider a more general metric with the mass of black hole is time $\left( t \right)
$ and radius $\left( r \right)
$ dependent. As generally known, this subject is well described by the Vaidya metric. To get an exact $\left( t-r \right)
$ dependence of the Hawking temperature, we insist to work in using Schwarzschild like metric for the Vaidya space time \cite{17} rather than standard Eddington-Finkelstein metric which is widely used for this case. We derive the general form for Hawking temperature both in the radial null geodesic and complex paths methods. We arrive only at the general form since one has to get the exact $\left( {t - r} \right)$ dependence of mass which of course needs a specific model of the related black hole. Until now, even though some models for dynamical black holes had been proposed, e.g. Refs. \onlinecite{23} and \onlinecite{24}, but efforts to get better one are still performed.\newline
The organization of our paper is as follows. In the second section, we will derive the Hawking temperature for a metric with a varying mass by the use of radial null geodesic method. In the third section, the Hawking temperature is obtained by the complex paths method, by considering only the lowest order expansion of action. At the fourth section, we show the possible expression of entropy. In the last section, we give a conclusion for our work. For the rest of this paper, we use the unit dimension: Newton constant, light velocity in vacuum, and Boltzman constant, $G = c = k_B  = 1$.

\section{Radial Null Geodesic Method}
Previously, a tightly connected work has been performed by T. Clifton \cite{8}, but with different method of calculations. In this work, we keep that the metric contains the mass of black hole as the function of standard radius-time $\left( r-t \right)
$ coordinate rather than Eddington-Finkelstein radiation coordinate $\left( u-v \right)
$. We start with the metric derived by Farley and D'Eath \cite{17} for Vaidya space-time
\begin{eqnarray}
	ds^2  =  - \left( {\frac{{\dot m}}{{x\left( m \right)}}} \right)^2 \left( {1 - \frac{{2m}}{r}} \right)dt^2  + \left( {1 - \frac{{2m}}{r}} \right)^{ - 1} dr^2  + r^2 d\Omega ^2.\label{eq:1} 
\end{eqnarray}
In the above, the black hole mass $m$ varies with time $t$ and radius $r$, $m \equiv m\left( {r,t} \right)
$. The $x\left( m \right)
$ is an arbitrary function of mass and $d\Omega ^2  = d\theta ^2  + \sin ^2 \theta d\varphi ^2 
$ is the metric of unit $2-$sphere. We will let the arbitrariness of $x\left( m \right)
$ since the determination of this function depends on the model of corresponding black hole. A short discussion of the model will be given at the last section but will not the main subject of this paper. The metric (\ref{eq:1}) has the general form
\begin{eqnarray}
	ds^2  =  - F\left( {r,t} \right)dt^2  + G\left( {r,t} \right)^{ - 1} dr^2  + r^2 d\Omega ^2,\label{eq:2}
\end{eqnarray}
where in our case $F\left( {r,t} \right) \equiv \dot m^2 x\left( m \right)^{ - 2} \left( {1 - 2mr^{ - 1} } \right)
$ and $G\left( {r,t} \right) \equiv 1 - 2mr^{ - 1} 
$. The general form (\ref{eq:2}) will simplify our next calculation. Unless for specific purposes, we will write $x\left( m \right)
$, $F\left( {r,t} \right)
$, and $G\left( {r,t} \right)
$ as $x$, $F$, and $G$ respectively for the sake of brevity.\newline
It turns out that the metric (\ref{eq:1}) has a coordinate singularity at $r_h  = 2m
$ which of course is time dependent. Painleve transformation that is used to remove the coordinate singularity for a metric with time-like Killing vector, is also applicable for this analysis. By transforming
\begin{eqnarray}
	dt \to dt - \sqrt {\frac{{1 - G}}{{FG}}} dr\label{eq:3}
\end{eqnarray}
the metric (\ref{eq:2}) changes into
\begin{eqnarray}
	ds^2  =  - F\left( {r,t} \right)dt^2  + 2F\sqrt{\frac{{1 - G}}{{FG}}} dtdr + dr^2  + r^2 d\Omega ^2,\label{eq:4}
\end{eqnarray}
and therefore no coordinate singularity is found. Thus, by such a Painleve transformation, it is understandable that, in principle, a coordinate singularity in general relativity can be removed only by changing coordinate, without defining new physical condition or theorem. By this thought, it would be alright if we still insist in using the metric (\ref{eq:1}) for carrying out a tunneling process. In the null geodesic condition for $(1+1)$ dimension, $ds^2  = d\Omega ^2  = 0
$, the radial null geodesic can be obtained from (\ref{eq:4}) as
\begin{eqnarray}
	\frac{{dr}}{{dt}} = \sqrt {\frac{F}{G}} \left( { \pm 1 - \sqrt {1 - G} } \right),\label{eq:5}
\end{eqnarray}
where $ + ( - )
$ signs denote outgoing(ingoing) radial null geodesics.\newline
Near the horizon, we can expand the coefficient $F$ and $G$ by the use of Taylor expansion. Since both $F$ and $G$ are $\left( {t - r} \right)
$ dependent, and we only need their approximation values for short distances from a point (horizon), we could apply the Taylor expansion at a fixed time. So, we can write
\begin{eqnarray}
	\left. {F\left( {r,t} \right)} \right|_t  \simeq \left. {F'\left( {r,t} \right)} \right|_t \left( {r - r_h } \right) + \left. {O\left( {\left( {r - r_h } \right)^2 } \right)} \right|_t ,\label{eq:6}
\end{eqnarray}
and
\begin{eqnarray}
	\left. {G\left( {r,t} \right)} \right|_t  \simeq \left. {G'\left( {r_h ,t} \right)} \right|_t \left( {r - r_h } \right) + \left. {O\left( {\left( {r - r_h } \right)^2 } \right)} \right|_t ,\label{eq:7}
\end{eqnarray}
By the approximations (\ref{eq:6}) and (\ref{eq:7}) above, the dependence of radius to time in (\ref{eq:5}) can be approached by
\begin{eqnarray}
	\frac{{dr}}{{dt}} \simeq \frac{1}{2}\sqrt {F'\left( {r_h ,t} \right)G'\left( {r_h ,t} \right)} \left( {r - r_h } \right).\label{eq:8}
\end{eqnarray}
Now, we discuss the action of outgoing particle through the horizon. In the original work by Parikh and Wilczek \cite{4}, the imaginary action is written as
\[
{\mathop{\rm Im}\nolimits} S = {\mathop{\rm Im}\nolimits} \int\limits_{r_{{\rm{in}}} }^{r_{{\rm{out}}} } {p_r dr}  = {\mathop{\rm Im}\nolimits} \int\limits_{r_{{\rm{in}}} }^{r_{{\rm{out}}} } {\int\limits_0^{p_r } {dp_r 'dr} } 
\]
\begin{eqnarray}
	 = {\mathop{\rm Im}\nolimits} \int\limits_{r_{{\rm{in}}} }^{r_{{\rm{out}}} } {\int\limits_0^H {\frac{{dH'}}{{{\textstyle{{dr}\over{dt}}}}}} } dr.\label{eq:9}
\end{eqnarray}
The above expression is due to the Hamilton equation $dr/dt = dH/dp_r |_r 
$ where $r$ and $p_r$ are canonical variables (in this case, the radial component of the radius and the momentum). As a reminder, the action of a tunneled particle in a potential barrier higher than the energy of the particle itself will be imaginary, $p_r  = \sqrt {2m\left( {E - V} \right)} 
$. Different from discussions of several authors for a static black hole mass, e.g. Refs. \onlinecite{4}, \onlinecite{8}--\onlinecite{15}, \onlinecite{21} and \onlinecite{22}, the outgoing particle's energy must be time dependent for black holes with varying mass. So, the $dH'
$ integration at (\ref{eq:9}) is for all values of outgoing particle's energy, say from zero to $ + E\left( t \right)
$.\newline
By using approximation (\ref{eq:8}), we can perform the integration (\ref{eq:9}). For $dr$ integration, we can perform a contour integration for upper half complex plane to avoid the coordinate singularity $r_h$. The result is
\begin{eqnarray}
	{\mathop{\rm Im}\nolimits} S = \frac{{2\pi E\left( t \right)}}{{\sqrt {F'\left( {r_h ,t} \right)G'\left( {r_h ,t} \right)} }}.\label{eq:10}
\end{eqnarray}
Since the tunneling probability is given by $\Gamma  \sim \exp \left[ { - {\textstyle{2 \over \hbar }}{\mathop{\rm Im}\nolimits} S} \right]
$, equalizing it with the Boltzmann factor $\exp \left[ { - \beta E\left( t \right)} \right]
$ for a system with time dependent energy we obtain 
\begin{eqnarray}
	T_H  = \frac{{\hbar \sqrt {F'\left( {r_h ,t} \right)G'\left( {r_h ,t} \right)} }}{{4\pi }}.\label{eq:11}
\end{eqnarray}
As seen in expression (\ref{eq:11}), the Hawking temperature $T_H$ is time dependent. We will see it later that it is also radius dependent. Partial derivatives of $F$ and $G$ with respect to $r$ are easily to be found in the following forms
\[
F' = \frac{{2m}}{{x^2 }}\left[ {m'\left( {1 - \frac{{2m}}{r}} \right) - \frac{{mx'}}{x}\left( {1 - \frac{{2m}}{r}} \right) + m\left( { - \frac{{m'}}{r} + \frac{m}{{r^2 }}} \right)} \right],
\]
and
\[
G' = \frac{2}{r}\left[ { - m' + \frac{m}{r}} \right].
\]
Inserting the values of $r_h  = 2m
$ to the last expressions yields Hawking temperature (\ref{eq:11}) in a general form
\begin{eqnarray}
	T_H  = \frac{{\hbar \left( {{1 \mathord{\left/
 {\vphantom {1 2}} \right.
 \kern-\nulldelimiterspace} 2} - m'\left( {r_h ,t} \right)} \right)}}{{4\pi x\left( {r_h ,t} \right)}}. \label{eq:12}
\end{eqnarray}
In the above, the $m'$ and $x$ are written has the complete expression as $m'\left( {r_h ,t} \right)
$ and $x\left( {r_h ,t} \right)
$. These mean that both $m'$ and $x$ are evaluated at $r_h$ (the radius of event horizon). Our finding in (\ref{eq:12}) behaves as it is expected for radius independent black hole, where it means that $m'(r_h ,t) = 0
$ and we get $T_H  = {\hbar  \mathord{\left/
 {\vphantom {\hbar  {4\pi x\left( t \right)}}} \right.
 \kern-\nulldelimiterspace} {4\pi x\left( t \right)}}
$ which is close to an expression that had been guessed by Ashtekar long time ago, $T_H  = {\hbar  \mathord{\left/
 {\vphantom {\hbar  {8\pi m\left( t \right)}}} \right.
 \kern-\nulldelimiterspace} {8\pi m\left( t \right)}}
$.

\section{Complex Paths Method}
Massless scalar particles under the gravitational background $g_{\mu \nu } 
$ obey the Klein-Gordon equation
\begin{eqnarray}
	\frac{{ - \hbar ^2 }}{{\sqrt { - g} }}\partial _\mu  \left[ {g^{\mu \nu } \sqrt { - g} \partial _\nu  } \right]\phi  = 0.\label{eq:13}
\end{eqnarray}
For spherical symmetric black hole, we may reduce our attention only to $\left( {t - r} \right)
$ sector in the space-time, or in other words, we reduce to two dimensional black hole problems. Later, since we are dealing with massless particle described in (\ref{eq:13}), we can employ the radial null geodesic without Painleve transformation\footnote{For static case, it was shown in Ref. \cite{15} that the results with and without Painleve transformation are equal.}
${{dr} \mathord{\left/
 {\vphantom {{dr} {dt}}} \right.
 \kern-\nulldelimiterspace} {dt}} =  \pm \sqrt {FG} 
$ for outgoing and ingoing particle's path. Equation (\ref{eq:13}) under the background metric (\ref{eq:2}) simplifies to
\begin{eqnarray}
	\partial _t ^2 \phi  - \frac{1}{{2FG}}\left( {\dot FG + \dot GF} \right)\partial _t \phi  - \frac{1}{2}\left( {F'G + FG'} \right)\partial _r \phi  - fg\partial _r ^2 \phi  = 0.\label{eq:14}
\end{eqnarray}
By the standard ansatz for scalar wave function $\phi \left( {r,t} \right) = \exp \left[ { - {\textstyle{i \over \hbar }}S\left( {r,t} \right)} \right]
$, equation (\ref{eq:14}) leads to the equation for the action $S\left( {r,t} \right)
$
\[
\left( {\frac{{ - i}}{\hbar }\left( {\frac{{\partial ^2 S}}{{\partial t^2 }}} \right)} \right) - \frac{1}{{\hbar ^2 }}\left( {\frac{{\partial S}}{{\partial t}}} \right)^2  - \frac{1}{{2FG}}\left( {\dot FG + \dot GF} \right)\left( {\frac{{ - i}}{\hbar }} \right)\left( {\frac{{\partial S}}{{\partial t}}} \right)
\]
\begin{eqnarray}
	 - \frac{1}{2}\left( {F'g + fG'} \right)\left( {\frac{{ - i}}{\hbar }} \right)\left( {\frac{{\partial S}}{{\partial r}}} \right) - fg\left( {\frac{{ - i}}{\hbar }\left( {\frac{{\partial ^2 S}}{{\partial r^2 }}} \right) - \frac{1}{{\hbar ^2 }}\left( {\frac{{\partial S}}{{\partial r}}} \right)^2 } \right) = 0.\label{eq:15}
\end{eqnarray}
Now, our next step is to solve this equation. An approximation method can be applied by expanding the action in the order of Planck constant power,
\begin{eqnarray}
	S\left( {r,t} \right) = S_0 \left( {r,t} \right) + \sum\nolimits_n {\alpha _n \hbar ^n S_n \left( {r,t} \right)},\label{eq:16}
\end{eqnarray}
for $n=1,2,3,...$. The constant $\alpha_n$ is set to keep all the expansion terms have the action's dimension. Taking unit dimensions $G = c = k_B  = 1
$, $\alpha_n$ would have the dimension of $[m]^{ - 2n} 
$ which $m$ refers to the mass. It is clear that this expansion would lead to a very long equation. Due to the very small value of the Planck constant, many authors \cite{8,10,21} neglect the terms for $n \ge 1
$. This consideration is acceptable, and including higher terms is just adding correction for semi-classical derivation of Hawking temperature. By grouping all the terms into the same powers of $\hbar 
$, and take the term with lowest order of Planck's constant (zero power), one can write
\begin{eqnarray}
	FG\left( {\partial _r S_0 } \right)^2  - \left( {\partial _t S_0 } \right)^2  = 0,\label{eq:17}
\end{eqnarray}
Our next task is to find the solution for the last equation which is not too simple since our $FG$ is $(r-t)$ dependent. In the standard Hamilton-Jacobi method, $S_0 \left( {r,t} \right)
$ can be written into two parts, the time part which has the form of $Et$ and the radius part $\tilde S_0 \left( r \right)
$ which is in general a radius dependent only. Since our metric coefficients are both radius and time dependent, the standard method would not be applicable. We could generalized the method by making an ansatz
\begin{eqnarray}
	S_0 \left( {r,t} \right) = \int\limits_0^t {E\left( {t'} \right)dt'}  + \tilde S_0 \left( {r,t} \right).\label{eq:18}
\end{eqnarray}
At the first sight, it seems that the ansatz is rather strange, that is $S\left( {r,t} \right)
$ and $\tilde S_0 \left( {r,t} \right)
$ are both $t$ and $r$ dependent. The term $\int {E\left( {t'} \right)dt'} 
$ is more understandable, since the emitted particle's energy is continuum and time dependent. Let see how it works.\newline
From (\ref{eq:18}), one can write that 
\begin{eqnarray}
	\partial _t S_0 \left( {r,t} \right) = E\left( t \right) + \partial _t \tilde S_0 \left( {r,t} \right)\label{eq:19}
\end{eqnarray}
and
\begin{eqnarray}
	\partial _r S_0 \left( {r,t} \right) = \partial _r \tilde S_0 \left( {r,t} \right).\label{eq:20}
\end{eqnarray}
Since $\tilde S_0 \left( {r,t} \right)
$ is $t$ and $r$ dependent, one can write
\begin{eqnarray}
	\frac{{d\tilde S_0 \left( {r,t} \right)}}{{dr}} = \partial _r \tilde S_0 \left( {r,t} \right) + \partial _t \tilde S_0 \left( {r,t} \right)\frac{{dt}}{{dr}}.\label{eq:21}
\end{eqnarray}
Eliminating $dt/dr$ by the use of ${{dr} \mathord{\left/
 {\vphantom {{dr} {dt}}} \right.
 \kern-\nulldelimiterspace} {dt}} =  \pm \sqrt {FG} 
$, equation (\ref{eq:21}) can be written as
\begin{eqnarray}
	\frac{{d\tilde S_0 \left( {r,t} \right)}}{{dr}} \mp \frac{1}{{\sqrt {FG} }}\partial _t \tilde S_0 \left( {r,t} \right) = \partial _r \tilde S_0 \left( {r,t} \right).\label{eq:22}
\end{eqnarray}
Combining the first equation of (\ref{eq:17}), with equations (\ref{eq:18}) and (\ref{eq:22}) where we should note that the action equation
$- \left( {FG} \right)^{ - 1/2} \partial _t S_0 \left( {r,t} \right) = \partial _r S_0 \left( {r,t} \right)
$ is belong to outgoing particle and with the radial evolution is $dr/dt = \sqrt {FG} 
$, then we could write
\begin{eqnarray}
	 \mp \left( {FG} \right)^{ - 1/2} \left( {E\left( t \right) + \partial _t \tilde S_0 \left( {r,t} \right)} \right) = \frac{{d\tilde S_0 \left( {r,t} \right)}}{{dr}} \mp \left( {FG} \right)^{ - 1/2} \partial _t \tilde S_0 \left( {r,t} \right).\label{eq:23}
\end{eqnarray}
From (\ref{eq:23}), we can get the exact differentiation of $\tilde S_0 \left( {r,t} \right)
$
\begin{eqnarray}
	\frac{{d\tilde S_0 \left( {r,t} \right)}}{{dr}} =  \mp \left( {FG} \right)^{ - 1/2} E\left( t \right),\label{eq:24}
\end{eqnarray}
and the solution of $\tilde S_0 \left( {r,t} \right)
$ can be obtained by integration
\begin{eqnarray}
	\tilde S_0 \left( {r,t} \right) =  \mp E\left( t \right)\int {\frac{{dr}}{{\sqrt {FG} }}}.\label{eq:25} 
\end{eqnarray}
The integration (25) can be evaluated by adopting the value of $\int {\left( {FG} \right)^{ - 1/2} dr} 
$ as in obtaining expression (\ref{eq:10}) from (\ref{eq:9}) along with it's approximation method (near horizon Taylor expansion). The result for the integration (\ref{eq:25}) is
\begin{eqnarray}
	\tilde S_0 \left( {r,t} \right) =  \mp E\left( t \right)\frac{{i\pi }}{{\sqrt {F'G'} }}.\label{eq:26}
\end{eqnarray}
The equation gives the complete action
\begin{eqnarray}
	S\left( {r,t} \right) = \int\limits_0^t {E\left( {t'} \right)dt'}  \mp E\left( t \right)\frac{{i\pi }}{{\sqrt {F'G'} }}.\label{eq:27}
\end{eqnarray}
The signs $+ \left(  -  \right)
$ in expression (\ref{eq:27}) refer to the action for ingoing (outgoing) particle.\newline
Back to our first ansatz for scalar wave function, $\phi  = \exp \left[ {{\textstyle{{ - i} \over \hbar }}S\left( {r,t} \right)} \right]
$, the wave function for ingoing and outgoing massless scalar particle can be read of as
\begin{eqnarray}
	\phi _{in} \left( {r,t} \right) = \exp \left[ { - \frac{i}{\hbar }\left( {\int\limits_0^t {E\left( {t'} \right)dt'}  + E\left( t \right)\frac{{i\pi }}{{\sqrt {F'G'} }}} \right)} \right]\label{eq:28}
\end{eqnarray}
and
\begin{eqnarray}
	\phi _{out} \left( {r,t} \right) = \exp \left[ { - \frac{i}{\hbar }\left( {\int\limits_0^t {E\left( {t'} \right)dt'}  - E\left( t \right)\frac{{i\pi }}{{\sqrt {F'G'} }}} \right)} \right]\label{eq:29}
\end{eqnarray}
respectively. Consequently, from (\ref{eq:28}) one can get the ingoing probability of particle as below
\begin{eqnarray}
	P_{in}  = \exp \left[ {\frac{2}{\hbar }\left( {{\mathop{\rm Im}\nolimits} \int\limits_0^t {E\left( {t'} \right)dt'}  + \frac{{\pi E\left( t \right)}}{{\sqrt {F'G'} }}} \right)} \right].\label{eq:30}
\end{eqnarray}
This ingoing probability must be equal to unity since all particles including the massless one are absorbed by the black hole. This consideration gives us the relation
\[
{\mathop{\rm Im}\nolimits} \int\limits_0^t {E\left( {t'} \right)dt'}  =  - \frac{{\pi E\left( t \right)}}{{\sqrt {F'G'} }},
\]
which leads the outgoing probability
\begin{eqnarray}
	P_{out}  = \exp \left[ { - \frac{{4\pi E\left( t \right)}}{{\hbar \sqrt {F'G'} }}} \right].\label{eq:31}
\end{eqnarray}
Finally, to get the Hawking temperature from the outgoing probability (\ref{eq:31}), we equate this probability expression with
$\exp \left[ { - \beta E\left( t \right)} \right]
$ which in Refs. \onlinecite{6} and \onlinecite{7} is called 'detailed balance' principle. It yields
\begin{eqnarray}
	T_H  = \frac{{\hbar \sqrt {F'\left( {r_h ,t} \right)G'\left( {r_h ,t} \right)} }}{{4\pi }}.\label{eq:32}
\end{eqnarray}
\section{General Form of Entropy}
The law of black hole mechanics which expresses the conservation of energy by relating the change in black hole mass m to the changes of its entropy $S_{bh} 
$, angular momentum $J$, and electric charge $Q$, is given by \cite{25}
\begin{eqnarray}
	dm = T_H dS_{bh}  + \Phi dQ + \Omega dJ
\end{eqnarray}
where $\Omega 
$ is the angular velocity and $\Phi 
$ is electrostatic potential. By simplifying the black hole with no angular velocity and electrical charges, one only has to pay attention on
\begin{eqnarray}
	dm = T_H dS_{bh}.\label{eq:33}
\end{eqnarray}
Inserting our previous result for $T_H$ from (\ref{eq:12}), we have an integral form for entropy as
\begin{eqnarray}
	S_{bh}  = \frac{{8\pi }}{\hbar }\int {\frac{{x\left( m \right)}}{{1 - 2m'}}dm}.\label{eq:34} 
\end{eqnarray}
To get an exact value of entropy which is intuitively should be equivalent to dynamical area of black hole, we must write $x\left( m \right)
$, $m'
$, and $\dot m
$ in their exact forms. This manner would be such an interesting work to be pursued further.
\section{Summary}
To summarize our findings above, we can state that we have derived a general expression for Hawking temperature (up to the model function $x\left( m \right)
$ and partial derivative of black hole mass with respect to radius) at (\ref{eq:11}) and (\ref{eq:12}) in two different semiclassical approaches, null geodesic and complex paths methods for a Vaidya black hole. This kind of black hole is regarded as the description of a more realistic one since it's time dependent decreasing mass due to the evaporation process. Clearly, the temperature would be time dependent as our findings. Both methods are found to give the same results. Then, we discuss he possible general form of discussed black hole entropy (\ref{eq:34}). This last finding perhaps can be compared to some models that had been proposed, e.g. Refs. \onlinecite{23} and \onlinecite{24}.

\section*{Acknowledgments}

H.M.S. is grateful to M. Siahaan and N. Sinurat for their supports and encouragements.

\end{document}